# METHOD OF ANALYSIS OF THE SPATIAL GALAXY DISTRIBUTION AT GIGAPARSEC SCALES. I. INITIAL PRINCIPLES

N.V.NABOKOV and Yu.V.BARYSHEV[1]

*Initial principles of a method of analysis of the luminous matter spatial distribution with sizes about thousands Mpc are presented. The method is based on an analysis of the photometric redshift distribution N(z) in the deep fields with large redshift bins Δz=0.1÷0.3. Number density fluctuations in the bins are conditioned by the Poisson's noise, the correlated structures and the systematic errors of the photo-z determination. The method includes covering of a sufficiently large region on the sky by a net of the deep multiband surveys with the sell size about 10°x10° where individual deep fields have angular size about 10'x10' and may be observed at telescopes having diameters 3 – 10 meters. The distributions of photo-z within each deep field will give information about the radial extension of the super large structures while a comparison of the individual radial distributions of the net of the deep fields will give information on the tangential extension of the super large structures. A necessary element of the method is an analysis of possible distortion effects related to the methodic of the photo-z determination.*

## 1. Introduction

Methods for observational evaluation of the sizes and contrast of inhomogeneities in the distribution of galaxies over a wide range of redshifts must be developed in order to construct a realistic model of the evolution of the large scale universe. Over the last two decades there has been a noticeable tendency in observations of the large scale structure of the universe toward the discovery of ever larger sizes of the structures which form galaxies and galactic clusters [1-3].

In the first wide angle survey of galactic redshifts in the 1980's (CfA) structures with a characteristic scale of 30 Mpc/h were found. For a long time these were regarded as the characteristic scale of the inhomogeneity of the universe [4,2]. However, in the deep "pencil" spectral surveys [5,6] structures with characteristic sizes of 130 Mpc/h were found. The existence of structures with characteristic sizes of hundreds of Mpc has also been confirmed by the recent wide angle spectral surveys 2dF and SDSS [3,7,8]. In these surveys, with a depth of $z < 0.6$, theoretically unexpected inhomogeneities with sizes on the order of 500 Mpc/h (e.g., the Sloan

[1] Institute of Astronomy , Saint Petersburg State University, Russia; e-mail: NabokovNikita@yandex.ru; yubaryshev@mail.ru

Great Wall [9]) were found. This suggests a need for direct observational limits on the possible presence of super large structures in the spatial distribution of galaxies, also at the large redshifts accessible through observations of deep fields.

Modern multiband deep surveys of galaxies [10], such as COMBO-17 [11], COSMOS [12], FDF [13], HUDF [14], ALHAMBRA [15], have opened up the possibility of direct observational determination of the sizes and contrast of super large inhomogeneities in the distribution of visible matter in the universe at redshifts of 0.5-5. These surveys contain $10^3$-$10^5$ galaxies with measured stellar magnitudes in several bands, so it is possible to study the radial distributions of galaxies based on photometric redshifts. The accuracy of the measurements of $z_{phot}$ depends on the method that is used and is usually 0.03(1 + z), which makes it possible to study scales greater than ~200 Mpc/h within the range of redshifts corresponding to the depth of the survey.

By using large redshift bins ($\Delta z = 0.1 \div 0.3$) containing large numbers of galaxies ($\Delta N \gtrsim 100$), the Poisson noise (~$1/\Delta N^{1/2}$) can be made small ($\sigma_p \lesssim 0.1$), so that fluctuations corresponding to super large inhomogeneities in the distribution of galaxies can be observed with a contrast greater than the Poisson level.

In this paper we propose a method for evaluating the sizes and contrast of super large inhomogeneities in the observed spatial distribution of galaxies on scales of thousands of Mpc. This method is based on an analysis of the deep fields found at the nodes of a grid covering a region of the celestial sphere large enough that it provides information on both the radial and tangential dimensions of inhomogeneities. As opposed to the observations of several deep fields proposed previously [6,15], where the individual fields are used only for analyzing the radial distributions, we propose using a grid of deep fields to search for and measure super large structures in both the radial and tangential directions. Thus, a method is being proposed for three dimensional tomography of the spatial distribution of visible matter in space on gigaparsec scales. This is needed in order to place observational limits on models of the evolution of the large scale structure of the universe and on the nature of the primordial density perturbations in the universe.

Section 2 is a description of the method and its basic components. Photometric redshifts and possible effects leading to distortions in the radial distribution of photo-z are discussed in Section 3. The main conclusions are given in Section 4.

## 2. Description of the method for volume tomography in space

A method of covering the celestial sphere with a grid of deep galactic surveys is proposed for discovering and measuring the parameters of super large structures within the visible distribution of matter in the universe. The optimum angular size for the grid of deep multiband surveys would be a cell of ~ 10°x10°, with deep fields of size ~ 10′x10′ located at its nodes. Then the distribution of the galaxies within each field would provide information on the extent of super large structures in the radial direction (using bins with Δz = 0.1-0.3), while comparison of the radial distributions in neighboring fields would provide information on the extent of the structures in the tangential direction with a step size of ~500 Mpc (10° at z ~ 1).

The proposed method of detecting possible super large structures involves the following stages:

- construction for each field of the observed distribution from photometric redshifts $\Delta N_{obs}(z, \Delta z)$ for the chosen bins $\Delta z$;
- construction of a theoretically expected distribution of redshifts $\Delta N_{obs}(z, \Delta z)$ for a uniform spatial distribution of galaxies in an artificial sample bounded by a fixed limiting visible stellar magnitude or for the entire volume of the sample;
- plotting of the relative deviations of the observed from the expected distributions in the fixed redshift bins;
- determining the regions where the observed fluctuations exceed the Poisson level $\sigma_p$ in the corresponding redshift bins and may, therefore, belong to correlated structures;
- determining the deviations in the radial redshift distributions for neighboring directions $(\alpha, \delta)$ in the sky.

**2.1. The scale lengths observationally accessible for a grid of deep fields.**

The metric distance in terms of the standard LCDM model is given by

$$r(z) = \frac{c}{H_0} \int_{\frac{1}{1+z}}^{1} \frac{dy}{y\sqrt{(\Omega_m^0/y + \Omega_v^0 y^2)}} \quad (1)$$

where c is the speed of light, $H_0 = 72$ km/s/Mpc, which corresponds to a normalized value of the Hubble constant $h = h_{100} = H_0/100 = 0.72$, the density parameters for matter and the vacuum are $\Omega_m^0 = 0.3$ and $\Omega_v^0 = 0.7$, and the age t(z) of the object (galaxy) is given by

$$t(z) = \frac{1}{H_0} \int_z^\infty \frac{dy}{(1+y)h(y)} \quad (2)$$

where

$$h(y) = \sqrt{\Omega_v^0 + \Omega_m^0(1+y)^3 - \Omega_k^0(1+y)^2} \quad (3)$$

The linear size of a region in the sky at time t which participates in expansion of space and corresponds to an observed angular size $\Delta\theta$ at time $t_0$ is given by

$$r_t(z) = \Delta\theta \times r(z), \tag{4}$$

where r(z) is given by Eq. (1).

The angular sizes of deep fields usually range from several minutes to one degree, so that the transverse dimensions of structures accessible to observations within a single field do not exceed a few tens of Mpc on a linear scale. However, for a grid of deep surveys the minimum dimensions of structures in the tangential direction correspond to the angular size of a cell and can range from hundreds to thousands of Mpc (Table 1).

TABLE 1. Linear dimensions Δr (Mpc) in the tangential direction corresponding to angular sizes Δθ (the angular size of a deep field or cell of the field grid) centered at z = 1 for the standard cosmological model (h = 0.72, $\Omega_v$ = 0.7, $\Omega_m$ = 0.3).

| Δθ      | 3′  | 10′ | $1^0$ | $10^0$ | $30^0$ | $60^0$ |
|---------|-----|-----|-------|--------|--------|--------|
| Δr(Mpc) | 2.8 | 9.3 | 55.7  | 557    | 1670   | 3341   |

The linear scale lengths which can be studied in the radial direction are determined by the bin size for the photometric redshift (Table 2) and are limited by the depth of the survey. The linear size for an interval Δz = 1 centered at z = 1 is 2392 Mpc.

TABLE 2. Linear dimensions Δr (Mpc) in the radial direction corresponding to intervals Δz (bins in the radial distribution) centered at z = 1 for the standard cosmological model (h = 0.72, $\Omega_v$ = 0.7, $\Omega_m$ = 0.3).

| Δz      | 0.1 | 0.2 | 0.3 | 0.4 | 0.5  | 1.0  |
|---------|-----|-----|-----|-----|------|------|
| Δr(Mpc) | 235 | 470 | 706 | 943 | 1180 | 2392 |

Because of the great depth of the surveys, in the radial direction it is possible to measure structural inhomogeneities on gigaparsec scales. The error in photo-z is on the order of $\sigma_z$ = 0.1 at z ~ 2, and this makes it possible to study structures with radial sizes in excess of 300 Mpc/h for redshifts of 0.5 - 5, where it is possible to evaluate the fluctuations in the number of galaxies in bins with Δz = 0.1 - 0.3.

**2.2. Expected distribution with respect to redshift**

For a uniform distribution of galaxies in space, a smooth distribution with respect to redshift is expected with possible fluctuations within the Poisson range for the relative error ($\sim 1/\Delta N^{1/2}$) corresponding to the number of galaxies in a Δz bin. Deep surveys are bounded in stellar magnitude by the samples, for which the distribution of galaxies over their redshifts is usually approximated by [17,18]

$$N_{\text{mod}}(z, \Delta z) = A z^{\alpha} e^{\left(-\frac{z}{z_0}\right)^{\beta}} \Delta z \quad (5)$$

where $\Delta N_{\text{mod}}(z, \Delta z)$ is the number of galaxies with redshifts within the interval $(z, z + \Delta z)$, the free parameters $\alpha$, $\beta$, $z_0$ and $z_0$ are found by least squares, and A is a normalization parameter corresponding to the condition. $\int N_{\text{mod}} dz = N_{\text{tot}}$. Equation (5) was also tested on model samples of distant galaxies bounded by a maximum visible stellar magnitude in which a uniform spatial distribution of galaxies with luminosities distributed according to a Schecther law was used [19].

When there are enough galaxies in the initial sample, it is also possible to construct complete subsamples, in terms of volume, within a limited range of z. For a uniform spatial distribution of galaxies the observed redshift distribution will be determined by the model for the evolution of the number and luminosity of the galaxies.

A second possibility for obtaining a theoretical distribution of the type (5) is numerical simulation of the distribution of galaxies in terms of the LCDM model. For example, a spatial distribution of galaxies on a light cone has been obtained [20,21], from which it is possible to extract the radial distribution of dark halos (regarded as galaxies) within a redshift range of 0 - 6. With that approach it is possible to obtain the form of the average distribution N(z), as well as the expected deviation from smooth behavior owing to correlated structures.

**2.3. Expected deviations from homogeneity.**

Deep surveys represent narrow conical cuts from the global spatial distribution of galaxies. For each redshift bin $(z, z + \Delta z)$ the theoretically expected dispersion in the relative deviations $\sigma^2$ is equal to the sum of the dispersions owing to correlated structures, $\sigma^2_{\text{corr}}$, and Poisson noise, $\sigma^2_{\text{p}}$ [22,23], i.e.,

$$\sigma^2(z, \Delta z) = \sigma^2_{corr} + \sigma^2_P \quad (6)$$

Poisson noise has a dispersion of

$$\sigma^2_P = \frac{\langle N^2 \rangle - \langle N \rangle^2}{\langle N \rangle^2} = \frac{1}{\langle N \rangle} \quad (7)$$

where the mean number $\langle N \rangle$ of galaxies within the volume corresponding to each bin $(z, z + \Delta z)$ can be found using Eq. (4), so that $\langle N \rangle = N_{\text{mod}}(z, \Delta z)$. Because of the sufficiently large number of observed galaxies in these redshift bins, $N_{mod}(z, \Delta z) \sim 100$, the contribution of Poisson noise to the observed fluctuations is limited to a small level of $\sigma_p \sim 0.1$.

According to Eq. (6), when deviations exceeding the level of $\sigma^2_p$ from the Poisson noise, we can speak of fluctuations in the number of galaxies within the corresponding bin owing to super large correlated structures. In this case, the expected dispersion (cosmic variance) $\sigma^2_{corr}$ owing to the structures can be found using the correlation function $\xi(r)$ according to the formula [22,23]

$$\sigma^2_{corr}(V) = \frac{1}{V^2} \int_V dV_1 \int_V dV_2 \xi(|r_1 - r_2|) \tag{8}$$

where $V = V(z, \Delta z)$ is the volume of integration corresponding to the bin $(z, z + \Delta z)$ under consideration.

As an approximate estimate for the expected dispersion $\sigma^2_{corr}$ in the case of a power law correlation function $\zeta(r) = \left(\frac{r_0}{r}\right)^\gamma$, the formula [22,24]

$$\sigma^2_{corr}(z, \Delta z) = \frac{J_2}{1+z}\left(\frac{r_0}{r_{eff}}\right)^\gamma \tag{9}$$

can be used, where $J_2 = \frac{72}{2^\gamma (3-\gamma)(4-\gamma)(6-\gamma)}$, $r_0$ is the correlation function parameter, and $1 + z$ is a factor that takes the growth in the inhomogeneity into account. For $\gamma = 1.8$, the constant $J_2 = 1.865$. The effective radius $r_{eff}$ of the spherical volume equivalent to the volume of a bin $(z, z + \Delta z)$ extended in the radial direction, is given by $r_{eff} = \left(\frac{3}{4\pi} r^2 \Delta r \Omega\right)^{1/3}$, where $r = r(z)$ and $\Delta r = \Delta r(\Delta z)$.

As measures of the deviation of the observed redshift distribution $N_{obs}(z, \Delta z)$ from the expected deviation $N_{mod}(z, \Delta z)$, for this bin $(z, z + \Delta z)$ we use the formula

$$\sigma_{obs}(z, \Delta z) = \frac{\Delta N_{obs}}{N_{mod}} = \frac{N_{obs}(z, \Delta z) - \langle N \rangle}{\langle N \rangle} \tag{10}$$

where the mean expected number of galaxies is given by Eq. (5). Based on Eq. (10), we distinguish regions with elevated (ODR_i or SLC_i) and reduced (UDR_i or SLV_i) densities of the number of galaxies relative to a Poisson level $\sigma_p$, i.e., regions with relative density fluctuations $\Delta N / N > + \sigma_p$ and $\Delta N / N < - \sigma_p$.

## 3. Evaluating photometric redshifts

**3.1. Photo-z in modern cosmology.**

The epoch of photometric redshift measurements is approaching in modern observational cosmology. Its main advantage over spectral measurements of z is that the distances to a large number of very faint galaxies can be estimated. In particular, it also offers a unique possibility for studying super large structures in the spatial distribution of galaxies.

In recent years the determination of the distances to faraway galaxies by means of photometric redshifts has become a generally accepted method in observational cosmology. Many papers have dealt with analyses of the accuracy of photo-z and compared it with spectral data [25-28].

The observational program ALHAMBRA [15] is in the process of completion. This program aims to measure the photo-z with an accuracy of 0.03(1+z) in 8 deep fields of size 40x40 angular minutes observed in 20 optical and 3 IR filters at the 3.5-m Calar Alto telescope. This will make it possible to study the evolution of both the distribution and the properties of the galaxies in radial directions from the observer. The term "cosmic tomography" has been introduced [15] for the study of the properties of the galaxies and structures in the radial direction, while the fields in the other directions are only needed for accumulating statistics of the radial properties. In this paper we propose using a grid of deep fields for studying super large structures in the tangential direction, as well as in the radial direction, which will make it possible to discover three dimensional structures.

Photometric redshifts have been used for calculating the power spectrum of the spatial distribution of LRG galaxies (depth z < 0.6) from the SDSS survey [29]. This made possible the discovery of a continuation of the power law growth in the fluctuations to wave numbers k = 0.005/(Mpc/h), corresponding to a scale length $\lambda = 2\pi / k = 1256$ Mpc. This indicates the existence of super large structures on gigaparsec scales.

Photo-z measurements have also been used to derive a distribution of dark matter in the COSMOS deep galaxy survey that turned out to be the same as the distribution of visible matter [12]. Strong fluctuations corresponding to known clusters of galaxies for z < 1.2 have been found [17] in a distribution of galaxies based on the photometric redshifts in the COSMOS survey. It has also been shown [30] that for large redshifts, real galactic clusters at z = 1.3 and 1.5 correspond to peaks in the distribution of galaxies based on photo-z in the COSMOS field.

**3.2. Accuracy of the photometric redshift method.**

Many analyses of the accuracy of photometric z measurements have been made [25-27,31,32]. The accuracy of the photo-z values depend on a number of factors:

- the number of filters and the wavelength interval that is covered;
- the photometric accuracy for each filter;
- the number and quality of the standard continuum spectra; and,
- the computer programs used to evaluate photo-z.

The number of filters used in various deep surveys varies widely, from 2 to 30, and covers a spectral range from the UV to the IR. Increasing the number of filters improves the accuracy of the photo-z estimates, but then the penetrance of the survey is reduced, since the passbands of the filters are smaller. The optimum number of filters has been analyzed by Benitez et al. [27].

For observations in the standard 5 filters (UBVRI), it has been found (Ref. 31, Table 2) that the accuracies of photo-z estimates ($\sigma_z$) for a photometric accuracy of $\Delta m = 0.1$ are 0.09 (z = 0.0- 0.4), 0.21 (z = 0.4- 1.0), 0.35 (z = 1.0- 2.0), 0.33 (z = 2.0- 3.0), 0.23 (z = 3.0- 5.0). For the optimum number of filters, 15-20, the accuracy of the estimates improves to $\sigma_z = 0.014(1+z)$ [27]. Adding IR filters also improves the number of crude errors (outliers) in the photo-z determinations. It is usually assumed that the error in determining photo-z is $\sigma_z = 0.03(1 + z)$, although projects are under way to attain an accuracy of 0.01 or even 0.003 [32]. The accuracy of spectral measurements of z is usually about 0.0001, or much better than that of photo-z, but these measurements cannot be performed for many faint galaxies.

There are a number of program packages for estimating photo-z which yield different accuracies [25]. These programs use different numbers of standards for the energy distribution in the continuum spectrum of the galaxies, as well as different mathematical criteria for fitting the spectra of the galaxies.

A key role in determining the photometric redshifts is played by the spectral standards for the energy distribution (SED). For example, the SED libraries of CWW [33] use the Hyperz program, which yields stable results even with a small number of filters for distant galaxies [31].

An important aspect of the search for super large structures is the fact that the requirements on the accuracy of the photo-z estimates are minimal. Since the elementary bins within which the galaxies are counted have sizes $\Delta z = 0.1$-0.3, the permissible accuracy of a photo-z measurement is at a level of $\sigma_z = 0.1 \div 0.3$.

**3.3. Selection effects and systematic errors.**

One important feature of the photometric redshift method is that it is based on the continuum energy distribution in the spectra of galaxies, so that it is independent of the visibility of spectral lines within different redshift intervals. Thus, for example, there is no "void" effect for the photo-z method in the region of $z \approx 2$ associated with the absence of visible lines; this has, for example, been invoked for interpreting a reduced number of spectrally observable gamma bursts in this range of redshifts [34-36].

However, there are some selection effects specific to the photometric redshift method owing to the visibility of certain details of the continuum spectra and a degeneracy in the solutions for a fixed system of filters in a given survey; for example, the visibility of the Balmer (3646 Å) and Lyman (912 Å) limits. This can lead to systematic errors and show up as inhomogeneities in the large scale radial distribution of galaxies.

The errors in measuring stellar magnitudes in different filters vary and depend on the individual features of observed galactic spectra, as well as on the accuracy with which the observations made at different angular resolutions are reduced (e.g., when combining optical and infrared data). In addition, the standards for the energy distribution in galactic spectra (SED) used in the photo-z method can differ from the actual continuum spectra. Thus, when the systematic effects are included, the formula for the observed dispersion in the fluctuations will have the form

$$\sigma^2(z,\Delta z) = \sigma_{corr}^2 + \sigma_P^2 + \sigma_{systematic}^2 , \qquad (11)$$

where the last term refers to the systematic effects of the method for evaluating photo-z.

Quantitative study of selection effects in the photometric redshift method is a complicated problem, upon the solution of which the reliability with which super large structures can be discovered in deep galactic surveys will depend. One possible way of studying the contribution of selection effects in the observed fluctuations $N_{obs}(z, \Delta z)$ in the numbers of galaxies is to model the procedure for measuring the photometric redshifts for an artificial catalog of uniformly distributed galaxies taking the features of each deep survey into account. In this method the model deviations $N_{mod}(z, \Delta z)$ can be subtracted from the observed deviations in the corresponding redshift bins and the excess in the fluctuations will be associated with an inhomogeneity in the radial distribution of the galaxies.

Another way of reducing the influence of selection effects could be to search for correlated deviations in the radial distributions of galaxies for a grid of deep surveys. When observations of a given field are made with different instruments, and different methods are used for evaluating photo-z, the distributions of the photometric redshifts within identical deep fields must coincide for real structures and be different when systematic distortions are present.

## 4. Conclusion

As an example of an evaluation of the resources required for carrying out a program of searching for super large structures by the method proposed here, let us consider observations made with the 6-m BTA telescope at the Special Astrophysical Observatory of the Russian Academy of Sciences. As a basis, let us take a program for the study of faint parent galaxies of identified gamma bursts, and galaxies in the deep field surrounding them. Observations with SCORPIO in a 4.3x4.3 angular minute field centered on the parent galaxy of the gamma burst GRB 021004 with exposures of 2600 s in the B band, 3600 s in V, 2700 s in R, and 1800 s in I yielded [37] the following limiting stellar magnitudes at a level of S/N > 3: 26.0 (B), 25.5 (V), 25.0 (R) and 24.5 (I). About 200 objects were identified with photometric z in the interval 0.1 - 4.

We can take a full night as a rough estimate of the observation time required for BVRI photometry of a single deep field. Thus, over 10 observation nights it is possible to cover a 20°x20° region of the sky for a grid cell size of 10°x10°, or 60°x60° for a 30°x30° cell size. Adding U and z filters does not change the order of magnitude of the observation time, but will increase the reliability of the photo-z determination.

The main difficulty in discovering super large structures in deep surveys of galaxies is taking quantitative account of the possible selection effects and systematic distortions in the distribution of galaxies according to the photometric redshifts associated with the technique for evaluating photo-z. Since this problem has not yet been solved, additional arguments such as spectral observations of distant galactic clusters will also play an important role.

In the next article in this series we discuss the application of the proposed method to a search for super large structures in available deep field data from COSMOS, FDF, HUDF, and HDF-N.

We thank a reviewer for useful comments that have greatly improved the presentation. We also acknowledge partial financial support from the Foundation for Leading Scientific Schools, grant No. NSh 1318.2008.2, and the Russian Foundation for Basic Research (RFFI), grant No. 09-02-00-143.